\documentclass[12pt,preprint]{aastex}
\usepackage{psfig}

\newcommand{\gae}{\mathrel{\raise .4ex\hbox{\rlap{$>$}\lower
  1.2ex\hbox{$\sim$}}}}

\slugcomment{Accepted for publication in the Astrophysical Journal}
\shorttitle{X-ray Image of M 87 Jet}
\shortauthors{Marshall et al.}

\begin{document}

\title{A High Resolution X-ray Image of the Jet in M 87}

\author{H.L. Marshall\altaffilmark{1},
B.P. Miller\altaffilmark{1},
D.S. Davis\altaffilmark{1},
E.S. Perlman\altaffilmark{2,3},
M. Wise\altaffilmark{1},
C.R. Canizares\altaffilmark{1},
and D.E. Harris\altaffilmark{4}}
\altaffiltext{1}{Center for Space Research, Massachusetts Institute of
	Technology, 77 Massachusetts Ave., Cambridge, MA 02139}
\altaffiltext{2}{Dept. of Physics, University of Maryland-Baltimore
	County, 1000 Hilltop Circle, Baltimore, MD 21250}
\altaffiltext{3}{Dept. of Physics and Astronomy, Johns Hopkins University,
	3400 North Charles St, Baltimore, MD 21218}
\altaffiltext{4}{Harvard-Smithsonian Center for Astrophysics, 60
	Garden St., Cambridge, MA 02138}
\email{hermanm@space.mit.edu, brendan@mit.edu,
dsd@space.mit.edu, perlman@umbc.edu, wise@space.mit.edu,
crc@space.mit.edu, harris@head-cfa.harvard.edu}

\begin{abstract}
We present the first high resolution X-ray image of the 
jet in M 87 using the {\em Chandra} X-ray Observatory.  There is clear
structure in the jet and almost all of the optically bright knots are
detected individually.  The unresolved core is the brightest
X-ray feature but is only 2-3 times brighter than
knot A (12.3\arcsec\ from the core) and the inner knot HST-1
(1.0\arcsec\ from the core).  The X-ray and optical positions of
the knots are consistent at the 0.1\arcsec\ level but the X-ray
emission from the brightest knot (A) is marginally
upstream of the optical emission peak.
Detailed Gaussian fits to the X-ray
jet one-dimensional profile show distinct X-ray emission that is not
associated with specific optical features.  The X-ray/optical flux
ratio decreases systematically from the core and X-ray emission is not
clearly detected beyond 20\arcsec\ from the core.  The X-ray spectra of
the core and the two brightest knots, HST-1 and A1, are consistent with
a simple power law ($S_{\nu} \propto \nu^{-\alpha}$)
with  $\alpha = 1.46 \pm 0.05$, practically ruling
out inverse Compton models as the dominant X-ray emission mechanism.
The core flux is significantly larger than expected from an advective
accretion flow and the spectrum is much steeper, indicating that the
core emission may be due to synchrotron emission from a small scale jet.
The spectral energy distributions (SEDs) of the knots are well fit by
synchrotron models.  The spectral indices in the X-ray
band, however are comparable to that expected
in the Kardashev-Pacholczyk synchrotron model but are much
flatter than expected in the pitch angle
isotropization model of Jaffe and Perola.  The break frequencies
derived from both models drop by factors of $10-100$ with distance
from the core.

\end{abstract}


\section{Introduction}

The Chandra X-ray Observatory is now resolving
the X-ray spatial structure along jets of radio galaxies
and quasars.  The jets of Cen A \citep{kraft00},
3C 273 \citep{marshall01}, and PKS~0637-752 \citep{schwartz00} are just
three examples of how {\em Chandra}
can be used to image jets at low to moderate redshifts.
In the 3C 273 and PKS~0637-752 jets,the X-ray power dominates the
spectral energy distributions (SEDs) of some of the jet knots and
simple synchrotron and synchrotron self-Compton models can
be ruled out directly.
These quasars show superluminal motion in
the radio-bright cores on scales of $\sim 100$ pc, which supports
the possibility that the large scale jets at $\gae$ 50 kpc may also be
moving superluminally, as required by a model where the X-ray emission
results from inverse Compton scattering of the cosmic microwave
background photons, such as suggested by \cite{celotti01}.
There is no direct evidence, however, for superluminal motion on
such large scales.

The M 87 jet is $\gae$ 30\arcsec\ long and shows significant details in
the optical and radio bands that differ only at the $\sim$0.1\arcsec\ level
\citep{sbm96}.  Superluminal motion is observed upstream
of knot A \citep{biretta99} which is just under a kpc from the core.
Previous X-ray observations by \citet{schreier82}, \cite{bsh91},
and \citet{harris97} detected the jet and ascribed
most of its emission to knot A, the brightest knot at optical and
radio wavelengths, so the M 87 jet offers a chance to test the
background upscattering model.
\cite{bohringer01} used the {\em XMM-Newton} telescope to
resolve the core from knot A and show that their spectra were similar
but the spatial resolution of none of the previous X-ray telescopes,
including {\em XMM-Newton}, were sufficient to resolve the jet into
more than one knot.  The emission from the core is also of
considerable interest for modelling of a possible advective accretion
flow \citep{reynolds96,dimatteo00}.

We present the first X-ray images with sufficient resolution
to identify and measure the X-ray emission from many knots in the jet.
We then produce SEDs for each distinct knot in the jet
and obtain X-ray spectra for several of these.

\section{Observations and Analysis}

M 87 was observed with the {\em Chandra} High Energy Transmission
Grating Spectrometer (HETGS) on 2000 July 17-18 (JD - 2450000
= 1743.17-1743.64).  The exposure time was 38048 s.
The readout detector was the Advanced CCD Imaging Spectrometer
(ACIS), which was read out in the timed exposure mode
with 3.2 s frame times.
The dispersed spectra of the core and knots were statistically
poor and showed no apparent emission lines so we ignored them
for the purposes of this work.  All remaining analyses concerned
only the image from the HETGS zeroth order.  The zeroth order
images are not affected by pileup because of the low count
rates ($< 0.1$ count per frame)
so point sources should be nearly Gaussian with a one dimensional
standard deviation of about
0.32\arcsec\ (0.75\arcsec\ FWHM) \citep{marshall01}.
At a distance of 16 Mpc \citep{tonry91}, 1\arcsec\ is 78 pc.

\begin{figure}
\plotone{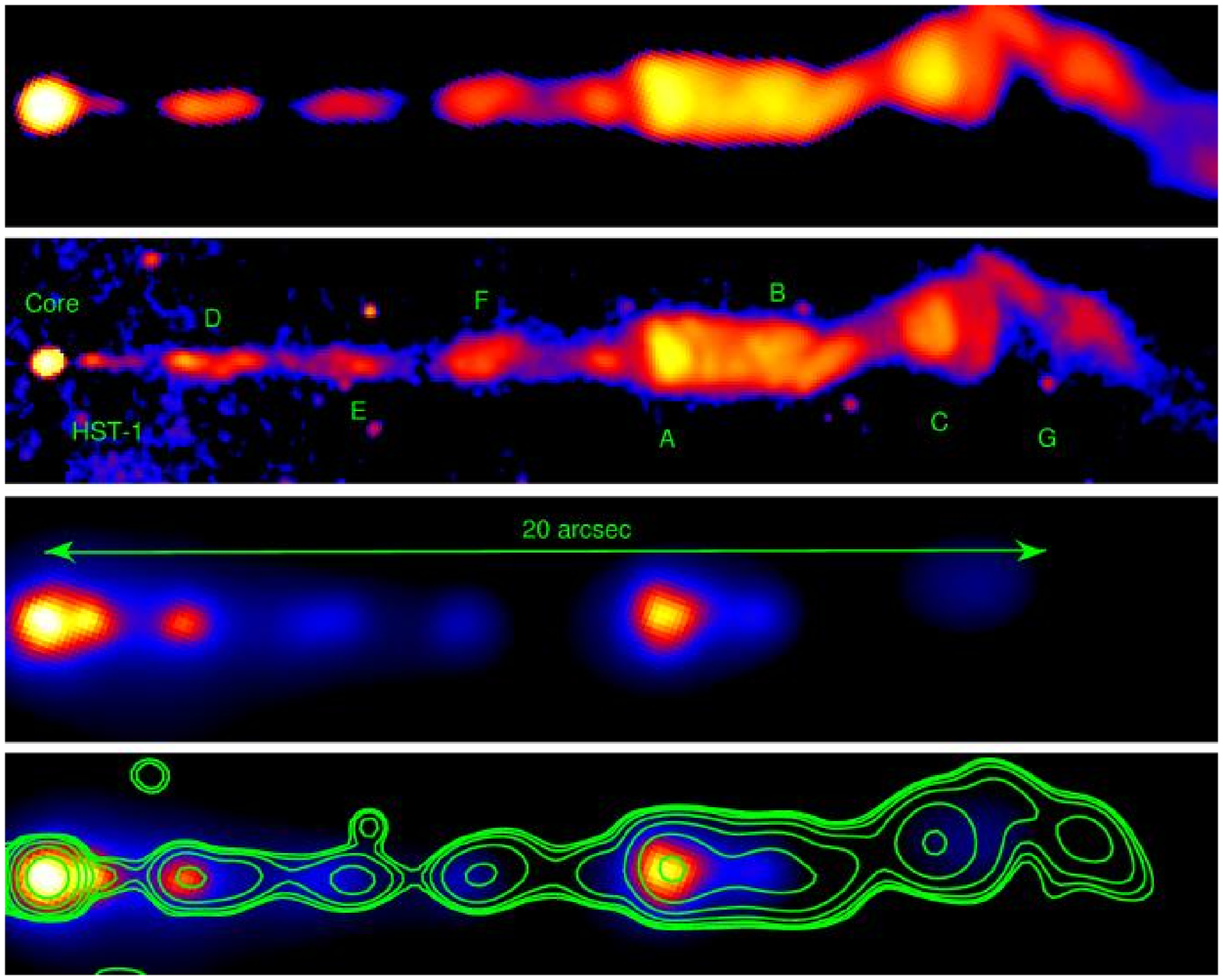}
\caption{Images of the jet in M 87 in three different bands,
rotated to be horizontal, and an overlay of optical
contours over the X-ray image.
{\em Top:} Image at 14.435 GHz using the VLA.  The spatial
resolution is about 0.2\arcsec.  
{\em Second panel:} The {\em Hubble} Space
Telescope Planetary Camera image in the
F814W filter from \citet{perlman01a}.
The brightest knots are labelled according to the nomenclature
used by \citet{perlman01a} and others.
{\em Third panel:} Adaptively smoothed {\em Chandra} image of the
X-ray emission from the jet of M 87 in 0.20\arcsec\ pixels.
The X-ray and optical images have been registered to each other
to about 0.05\arcsec\ using the position of the core.
{\em Fourth panel:} Smoothed
{\em Chandra} image overlaid with contours
of a Gaussian smoothed version of the HST image, designed
to match the {\em Chandra} point response function.
The X-ray and optical images have been registered to each other
to about 0.05\arcsec\ using the position of the core.
The HST and VLA images are displayed using a logarithmic
stretch to bring out faint features while the X-ray image
scaling is linear.}
\label{fig:image}
\end{figure}

\begin{figure}
\epsscale{0.8}
\plotone{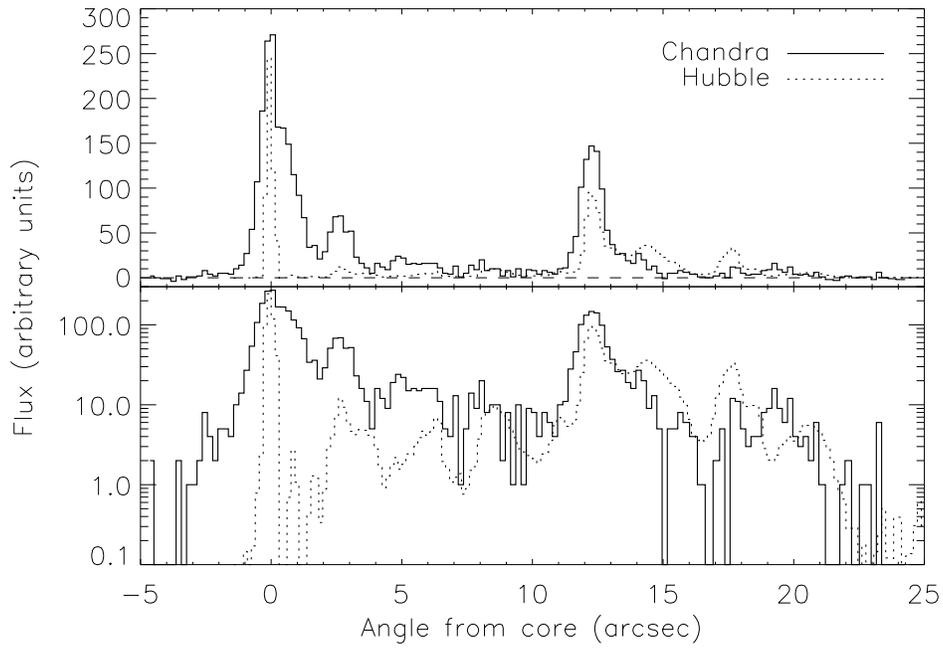}
\caption{Profiles of the M 87 jet in the X-ray (solid,
in counts per 0.2\arcsec\ bin) and optical bands
(dashed) using the images shown in Fig.~\ref{fig:image}.
The optical data were scaled
such that 28.4 nJy per 0.152\arcsec\ bin
corresponds to a vertical value of 200.
With logarithmic scaling (bottom panel), it is very clear
that the X-ray fluxes decrease relative to the optical
flux with increasing angle from the core.
\label{fig:profile} }
\end{figure}

Figure~\ref{fig:image} shows the X-ray image binned
in 0.2\arcsec\ pixels and adaptively smoothed.
For comparison, images from
the VLA and HST are also shown, taken from observations
reported by \citet{perlman01a}.
Qualitatively, the X-ray emission from the jet
is much brighter near the core compared to the optical
emission.  Knot HST-1 is the best example of the
difference, being the second brightest knot in the
X-ray image but it is the faintest of the knots optically.
By way of contrast, X-ray emission is barely
detectable beyond knot B, which is about 14\arcsec\ from the core.
The jet is nearly one-dimensional,
so a profile is used for quantitative analysis
(Fig.~\ref{fig:profile}),
derived by summing data in a 1.5\arcsec\ wide window at
a position angle of -70.4\arcdeg, which is defined by
the center of knot A.
The X-ray flux reaches the background level at 21\arcsec\ from
the nucleus.

Gaussians were fitted to the knots in the one dimensional profile of the
jet.  The results of the fits
are given in Table~\ref{tab-fluxes}.  There is definite X-ray emission
that is not included in any of the fitted regions, which were restricted
to the locations of optically emitting knots.  In particular,
there is significant X-ray emission between knots D and E that is
not included in either of these fitting regions which we
label ``DX''.  Similarly,
there is a ``bridge'' of X-ray emission between knots A and B,
which corresponds to a region downstream of knot A.  The X-ray
flux clearly drops more rapidly than the optical flux does in
this downstream region; the X-ray emission we label as knot G
is perhaps more closely associated with the downstream end
of knot C, based on the image (Fig.~\ref{fig:image}).

The distance from the core to the peak of the
X-ray emission of knot A is 12.34 $\pm$ 0.02\arcsec, which
is within 0.1\arcsec\ of the distance derived from the HST
data, 12.43 $\pm$ 0.01\arcsec\ (see
table~\ref{tab-fluxes}).  Previous estimates of
this separation using X-ray images with lower angular resolution
\citep{neumann97,bohringer01} gave smaller values, about 11.5\arcsec.
These estimates were probably biased by the flux in the bright
HST-1 and D knots.  Indeed, we find that the centroid of
the core, HST-1, and D knots is 11.66 $\pm$ 0.02\arcsec
from knot A, which
is consistent with the {\it ROSAT} and {\it XMM} results.
Thus, although we also find that the X-ray emission of knot A
is upstream of the peak optical emission by 0.09 $\pm$ 0.03\arcsec,
the offset is significantly smaller than previously estimated.
Other knots show less significant displacements.

We tested for cross-jet extent of knot A by fitting Gaussians
to the cross-jet profiles of the core/HST-1 knot combination
as well as to knot A itself.  The standard deviations
(FWHMs) are 0.371
$\pm$ 0.009\arcsec\ (0.872\arcsec) for the core region
and 0.439 $\pm$ 0.016\arcsec\  (1.032\arcsec) for knot A,
indicating that knot A is broader than the core.
Assuming that the core image represents a point source, we
estimate the intrinsic FWHM of knot A to be 0.55 $\pm$ 0.07\arcsec.
In the optical images, this knot is 1.0\arcsec\ wide
\citep{sbm96}, so we
conclude that the X-ray emission is more concentrated than the
optical emission from this knot in the cross-jet
direction following the trend observed by \citet{sbm96} that
the knot appears narrower in the UV than in the optical
and radio bands.

Four regions were selected for X-ray spectral fitting using
0.5-7.5 keV
data reprocessed by the Chandra X-ray Center using CIAO 2.0b
and the most recent response matrices for ACIS chip S3.
Results are given in Table~\ref{tab-spectra}.
The reduced $\chi^2$ values are all acceptable at the
95\% confidence level.  The spectral indices of all regions
are consistent with indices found by \cite{bohringer01},
who did not resolve knots HST-1 and D from the core.
These indices are consistent with a single
value of $\alpha$ of 1.46 $\pm$ 0.05.
We tested the pulse height spectra of most
knots against the spectrum of the core in order to determine
if the spectra changed shape.  The core was defined as a region
1\arcsec\ in diameter placed to exclude most of the flux of
the HST-1 knot.  Smirnov two-sample tests
showed that the spectra of all knots, including HST-1,
were consistent with
that of the core at the 90\% confidence level.
By contrast, the radio-optical spectral indices of these knots
are 0.66-0.71 and the optical spectral slopes are
in the range 0.65-0.90 \citep{perlman01a}.

Flux densities of several knot regions are given in
Table~\ref{tab-fluxes}, based on Gaussian fits to the
one dimensional profile shown in Fig.~\ref{fig:profile}.
We find that the total jet power is 1.59 $\pm$ 0.07
times that of the core power in the 0.5-5.0 keV band.
The unabsorbed 0.5-10 keV luminosity of the core is 4.4 $\times 10^{40}$
erg s$^{-1}$.

\section{Discussion}

The observed luminosity and
spectral index of the unresolved core can be used to
constrain models of the accretion emission.
\citet{reynolds96} estimated that
the X-ray flux, $\nu S_{\nu}$, was about 1.6 $\times 10^{-12}$
erg s$^{-1}$ cm$^{-2}$ at 1 keV and $< 7 \times 10^{-12}$
erg s$^{-1}$ cm$^{-2}$ at 5 keV at resolutions of 4\arcsec\ and
200\arcsec, respectively.  Reynolds et al.\ argued that the
expected X-ray flux from an advective
flow would be a factor of 10-100 smaller than the observed
values and suggested that the core optical and X-ray emission would
instead result from a jet.  We find somewhat smaller
fluxes of 6.0 and 3.0 $\times 10^{-13}$
erg s$^{-1}$ cm$^{-2}$ at 1 and 5 keV at a resolution of
0.5\arcsec; these fluxes are still $\sim 10\times$ larger
than the expected emission due to accretion.
It is also significant, however, that the advective flow model
proposed by \cite{reynolds96,dimatteo00} produce X-ray emission
by thermal bremsstrahlung at a characteristic temperature
of $2 \times 10^9$ K, giving an X-ray spectrum that would be
much flatter than observed.
The new X-ray observations indicate that the M 87 core is
probably dominated by a jet in the X-ray band.  This interpretation
is supported by 10 $\mu$ measurements \citep{perlman01b}.
It is important to remember, however,
that this conclusion is model-dependent, relying on modelling
of the emission from advective flows.
\citet{allen00} used ASCA spectra
to argue for a hard spectral component ($\alpha = 0.4$) which
they associate with the core.  \citet{gm99} also argued for
a hard component based on {\em BeppoSAX} data for the
inner 2\arcmin\ with a flux about 2$\times$ smaller
than found by \citet{allen00}.
The expected 5 keV flux in the \citet{allen00}
model is a factor of 20 larger than our measurement, which
indicates that the hard flux is extended, a conclusion
also reached by \cite{matsumoto98} in M 87 and
by \cite{loewenstein01} for other ellipticals modelled by \citet{allen00}.

\begin{figure}
\epsscale{0.7}
\plotone{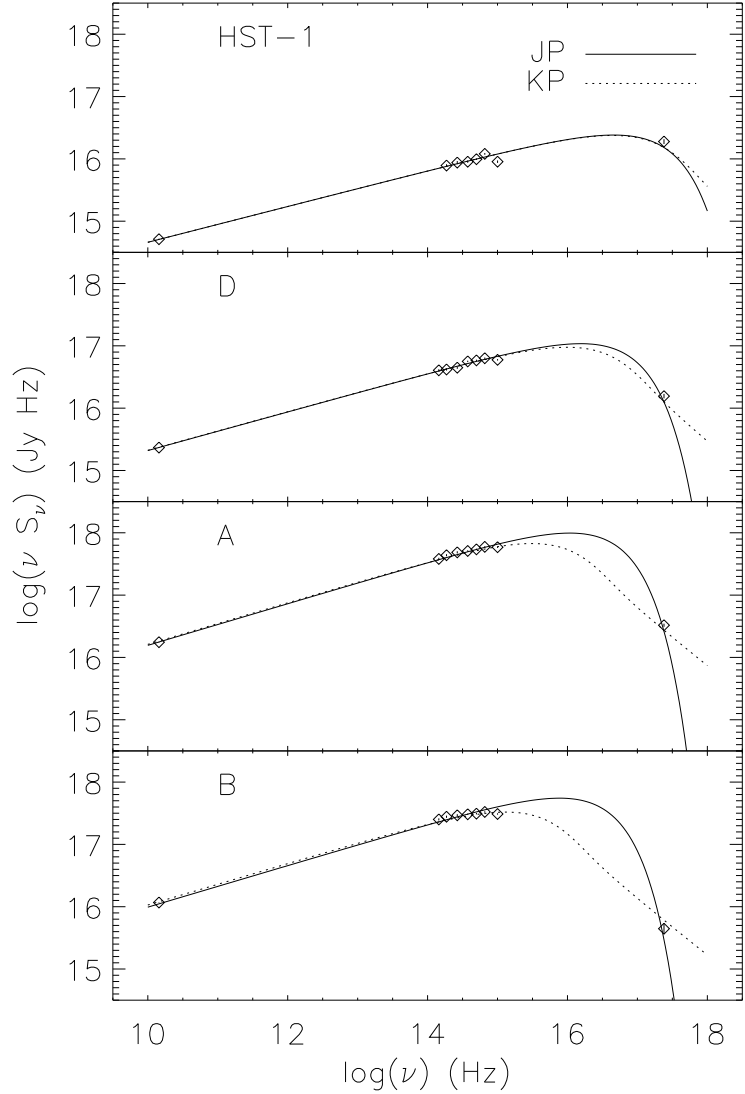}
\caption{The spectral energy distributions (SEDs) of the M 87
jet knots.  Error bars are given for each data point and are
usually much smaller than the plot symbols.
There is clear evidence of spectral steepening
in the optical-UV data for knots A and B but a stronger break or
cutoff is required to fit the X-ray flux density.
The data were fitted to
\cite{jp73} and Kardashev-Pacholczyk \cite{k62,p70}
synchrotron models, shown as dotted and
dashed lines, respectively.  The KP model can fit all
knot SEDs reasonably well while the JP model
fits the optical data of knot B poorly.
The break frequencies obtained for the outlying knots
are quite different for these
two models; see also Fig.~\ref{fig:synch-break}.
\label{fig:sed} }
\end{figure}

With the high resolution of {\em Chandra}, we have measured
the X-ray flux densities of the knots detected in HST images in
order to construct reliable spectral energy distributions (SEDs)
(Fig.~\ref{fig:sed}) using radio and optical
flux densities from \citet{perlman01a}.
We have applied the usual synchrotron formulae \citep{p70}
to the radio-optical portions of the SEDs for
knots HST-1, D, A, and B, finding
equipartition magnetic field values in the range
250 to 320 $\mu$G and synchrotron luminosities of
$10^{40-42}$ erg s$^{-1}$.  The particle Lorentz factors required to
produce X-rays in these fields are close to
$\gamma \approx 3 \times 10^7$ and the synchrotron loss lifetimes
are of order 3 to 10 years (for $10^{18}$ Hz and $10^{17}$ Hz, respectively).
\citet{bsh91} obtained similar results for their synchrotron models
and the lifetimes are consistent with observed
X-ray variability time scales \citep{harris97}.

Following \citet{perlman01a}, we fitted the SEDs
of each knot to the \cite{jp73}
and Kardashev-Pacholczyk \cite{k62,p70} synchrotron models,
labelled JP and KP, respectively.  These models are described
by Perlman et al.; briefly, the JP model involves isotropization
of the electron pitch-angle distribution while the KP model
does not.  Continuous injection models \citep{hm87} generally fail to
fit the X-ray data, as shown by Perlman et al.
They determined that this model might still apply to knot A
if the X-ray emission region were $\sim 5\times$ smaller than
the optical emission region.  We find that the width of the X-ray
emission from knot A is about half that of the optical emission,
so the emission volume could indeed be as small as required
for the continuous injection model.  These data are not
sufficient, however, to test for the extremely small
volumes required for this model to fit other knots.
Now that the X-ray flux density for knot HST-1 is known, we
find that it may also be effectively modelled with a
continuous injection model.

\begin{figure}
\epsscale{0.8}
\plotone{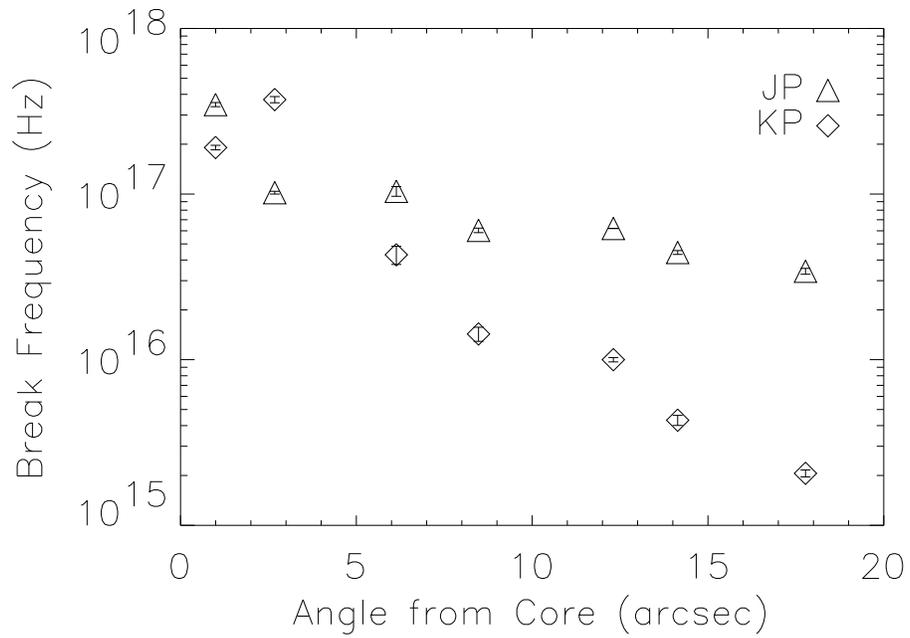}
\figcaption{
Synchrotron break frequency from fits to the spectral
energy distributions of each knot, as shown in
Fig.\ref{fig:sed}.  The JP and KP synchrotron models
are taken from \citet{jp73} and \cite{k62,p70},
respectively.  Note the systematic downward trend of
the break frequencies.
\label{fig:synch-break} }
\end{figure}

Fig.~\ref{fig:sed} shows that the KP model can fit all
knot SEDs reasonably well while the JP model
fits the optical data of knot B poorly.
The break frequencies are quite different for these two models
for knots furthest from the core.
Fig.~\ref{fig:synch-break} shows
that the break frequency, $\nu_B$, drops systematically with
position along the jet after knot HST-1.
We do not yet have a model that predicts this behavior.
The KP and JP models predict specific spectral indices in the
X-ray band.  For the KP model, the
spectral indices are nearly identical for all knots in
the jet: 1.7-2.1; these values are systematically larger than
the observed average of 1.46.
The JP model, however, would result in
X-ray spectral indices that are usually larger than 3.5,
which is much steeper than observed.  These two models
are used for illustrative purposes, giving an indication
of the difficulties encountered in global fitting which,
in turn suggests the need for spatially stratified
emission regions \cite{perlman99,perlman01a}.

Since inverse Compton (IC) models using synchrotron emission
as seed photons fail by factors of hundreds
\citep{neumann97,bohringer01}, either the field is significantly
smaller than the equipartition value \citep{hb97} or the seed
photons come from the cosmic microwave background (CMB),
which would be seen in
the jet frame to have an energy density augmented by the jet's Lorentz
factor ($\Gamma$) squared.  We have applied the formulation of
\citet{hw01}, who solve for the beaming
parameters under the assumption of a synchrotron source with
equipartition field.  While this sort of beaming model provides
acceptable solutions for some sources such as
PKS 0637-752 \citep{celotti01}, in the case of M 87 rather extreme
conditions would be required.  In particular, the angle of the jet to
the line of sight would have to be less than 5\arcdeg\ and the
beaming factors would be in the range 10 to 45.  Such models would
conflict with our current understanding of both VLBI scale and arcsec
scale studies which essentially concur that the angle to the line of
sight is of order 20\arcdeg\ and the Lorentz factors are of order a
few \citep{biretta99}.  Our synchrotron model fits into this conventional
interpretation; there may well be significant beaming in the kpc
jet of M87 but we are not in a favored position to see the primary
effects.  In all IC models, the X-ray spectral slopes should be
comparable to those in the radio band while we find that the X-ray
spectra are significantly steeper, even for knots HST-1 and D where
some beaming might be expected, based on the observation of superluminal
motion \citep{biretta95,biretta99}.  Thus, purely based on the X-ray spectral
indices, IC models do not readily explain the X-ray data.

At such high spatial resolution, we begin to see that modelling
will be complicated by comparing emission from regions that are
not strictly co-spatial so that spatially stratified emission
regions may be required.  \citet{perlman99} suggested this model
in order to reconcile the radio and optical polarization vectors.
Particle acceleration in the knots is required in order to replenish
the supply of X-ray emitting electrons so
the highest energy electrons will emit in the X-ray band and
have the shortest lifetimes while the lower energy electrons
can radiate further down the jet.
We observe X-ray emission upstream of knot E,
and that knot A appears closer to the core in the X-ray band than
in the optical or radio bands.
Our estimates of the synchrotron lifetimes of the X-ray
emitting electrons -- 3-10 yr --
are consistent with the marginally significant offset we observe
between the optical and X-ray positions of knot A: $\sim 8$ pc,
given uncertainties in the angle of the jet to the line of 
sight and the bulk motion of the jet.
For Cen A, which is 5$\times$
closer, there are much clearer offsets \citep{kraft00,kraft01},
resulting from particle diffusion and energy loss.
Similarly, we may be observing
lateral diffusion of shock-accelerated electrons in the cross-jet
leading edge of knot A.  The X-ray emission region would, again, be
significantly smaller than the optical emission region.  Higher
resolution X-ray observations of the M 87 jet may well show
morphological differences that will require somewhat more complex
spectral modelling.

\acknowledgments

We thank Ralph Kraft for communicating results from {\em Chandra}
observations of Cen A in advance of publication.
This research is funded in part by NASA contracts NAS8-38249,
and SAO SV1-61010.  ESP was funded by NASA grant NAG5-9997 and
HST grant GO-7866
and DEH was supported under grant NAS8-39073.

\clearpage

\clearpage

\begin{deluxetable}{lcccc}
\tablecolumns{5}
\tablewidth{0pc}
\tablecaption{Flux Densities and Positions\tablenotemark{a}~ of Knots in
	the M 87 Jet \label{tab-fluxes} }
\tablehead{\colhead{Feature} & \colhead{$\theta_x$}  & \colhead{$S_{\nu}$\tablenotemark{b}}
	& \colhead{$\theta_o$} & \colhead{$\theta_r$} \\
 & \colhead{(\arcsec)} & \colhead{(nJy)} & \colhead{(\arcsec)} & \colhead{(\arcsec)} }
\startdata
Core	&	0.00			&       245 $\pm$	10	&	0	&	0\\
HST-1 	& 	1.02	$\pm$	0.03	&	82	$\pm$	7 	&	0.92	&	1.16 \\
D\tablenotemark{c}	&	2.71	$\pm$	0.03	&	67.9	$\pm$	5.6 	& 2.79 & 3.09 \\
DX\tablenotemark{c}  &	4.93	$\pm$	0.05	&	21.1	$\pm$	3.5	& \nodata & \nodata\\
E	&	6.16    $\pm$	0.06 	&	15.4	$\pm$	3.1	& 6.15 & 5.98 \\
F	&	8.50    $\pm$	0.08	&	10.0 	$\pm$	2.6	& 8.65 & 8.71 \\
A	&	12.34   $\pm$	0.02	&	142 $\pm$	8	& 12.43 & 12.49 \\ 
B	&	14.16   $\pm$	0.06 	&	19.2	$\pm$	3.5	& 14.45 & 14.44 \\
C	&	17.80   $\pm$	0.10	&	7.2	$\pm$	2.5	& 17.70 & 17.69 \\
G\tablenotemark{d}	&	19.33   $\pm$	0.07	&	11.6	$\pm$	2.7 & \nodata & \nodata\\
\enddata
\tablenotetext{a}{Positions are given as angular distances
from the core in units of arcsec for the X-ray, optical,
and radio bands, as designated by the subscript on $\theta$.
The systematic uncertainties in the optical and radio values
are estimated to be about 0.01\arcsec.}
\tablenotetext{b}{Flux density at 1 keV assuming
a spectral index of 1.46, the uncertainty weighted average
of values from table~\ref{tab-spectra}.}
\tablenotetext{c}{The X-ray flux of knot D is spatially associated
with the optical emission of knot D-east, as defined
by \citet{perlman99} 
while Knot DX is not clearly associated with
any of the D subknots found in HST images.}
\tablenotetext{d}{The X-ray emission appears to be more
closely related to the end of knot C rather than knot G
based on the position angle.}
\end{deluxetable}

\begin{deluxetable}{lccc}
\tablecolumns{4}
\tablewidth{0pc}
\tablecaption{Power Law Spectral Fits for Knots in the
	M~87 Jet \label{tab-spectra} }
\tablehead{\colhead{Region} & \colhead{$\alpha$\tablenotemark{a}} &
	\colhead{$\chi^2$ (dof)\tablenotemark{b}} }
\startdata
Core \& HST-1 	&	1.47$\pm$0.08  & 1.19 (61)  \\
Knot HST-1 	&	1.29$\pm$0.14  & 0.93 (22)  \\
Knot D 	&	1.33$\pm$0.17  & 1.53 (22)  \\
Knot A	&	1.57$\pm$0.10  & 1.12 (32)	\\ 
\enddata
\tablenotetext{a}{Uncertainties are 90\% confidence values for 1
interesting parameter for
a power law of the form $S_{\nu} \propto \nu^{-\alpha}$, with
the column density of interstellar gas fixed to
$2.5 \times 10^{20}$ cm$^{-2}$}
\tablenotetext{b}{Reduced $\chi^2$ for the number degrees of freedom
given in parentheses}
\end{deluxetable}


\begin{thebibliography}{}
\bibitem[Allen, Di Matteo, \& Fabian(2000)]{allen00}
	Allen, S.~W., Di Matteo, T., \& Fabian, A.~C.\ 2000, \mnras, 311, 493
\bibitem[Biretta, Zhou, \& Owen(1995)]{biretta95}
	Biretta, J.\ A., Sparks, W.\ B., and Owen, F.\ N.\ 1995, \apj, 520, 621
\bibitem[Biretta, Sparks, \& Macchetto(1999)]{biretta99}
	Biretta, J.\ A., Sparks, W.\ B., and Macchetto, F.\ 1999, \apj, 520, 621
\bibitem[Biretta, Stern, \& Harris(1991)]{bsh91}
	Biretta, J.\ A., Stern, C.\ P., and Harris, D.\ E.\ 1991, \aj, 101, 1632
\bibitem[B\"ohringer et al.(2001)]{bohringer01}
	B\"ohringer et al. 2001, \aap, 365, L181
\bibitem[Celotti, et al.(2001)]{celotti01}
	Celotti, A., Ghisellini, G., and Chiaberge, M. 2001, \mnras, 321, L1.
\bibitem[Di Matteo et al.(2000)]{dimatteo00}
	Di Matteo, T.,  Quataert, E., Allen, S.\ W., Narayan, R., \&
	Fabian, A.\ C.\ 2000, \mnras, 311, 507
\bibitem[Guainazzi \& Molendi(1999)]{gm99}
	Guainazzi, M.~\& Molendi, S.\ 1999, \aap, 351, L19 
\bibitem[Harris et al.(1997)]{harris97}
	Harris, D.\ E., Biretta, J.\ A., and Junor, W.\ 1997, \mnras, 284, L21
\bibitem[Harris \& Krawczynski (2001)]{hw01}
	Harris, D.\ E.\ H.\ \& Krawczynski, H.\ 2001, \apj, submitted
\bibitem[Heavens \& Meisenheimer (1987)]{hm87}
	Heavens, A.\ \& Meisenheimer, K.\ 1987, \mnras, 225, 335
\bibitem[Heinz \& Begelman (1997)]{hb97}
	Heinz, S.\ \& Begelman, M.\ 1997, \apj, 490, 653
\bibitem[Jaffe \& Perola (1973)]{jp73}
	Jaffe, W.\ J.\ \& Perola, G.\ C.\ 1973, \aap, 26, 421
\bibitem[Kardashev (1962)]{k62}
	Kardashev, N.\ S.\ 1962, Soviet Astronomy -- AJ, 6, 317
\bibitem[Kraft et al.(2000)]{kraft00} Kraft, R.\ P.\ et al.\ 
	2000, \apjl, 531, L9
\bibitem[Kraft et al.(2001)]{kraft01} Kraft, R.\ P.\ et al.\ 
	2001, \apj, submitted
\bibitem[Loewenstein et al.(2001)]{loewenstein01}
	Loewenstein, M., Mushotzky, R.~F., Angelini, L., Arnaud,
	K.~A., \& Quataert, E.\ 2001, \apjl, 555, L21
\bibitem[Marshall et al.(2001)]{marshall01}
	Marshall et al. 2001, \apjl, 549, L167
\bibitem[Matsumoto(1998)]{matsumoto98}
	Matsumoto, H.\ 1998, Ph.D.\ thesis, Kyoto University
\bibitem[Neumann et al.(1997)]{neumann97}
	Neumann, M.\ Meisenheimer, K.\ R\"oser, H.-J., and
	Fink, H.\ H. 1997, \aap, 318, 383
\bibitem[Pacholczyk (1970)]{p70}
	Pacholczyk, A.\ G.\ 1970, Radio Astrophysics (San Francisco: Freeman)
\bibitem[Perlman et al.(2001a)]{perlman01a}
	Perlman, E.\ S., Biretta, J.\ A., Sparks, W.\ B.,
	Macchetto, F.\ D., and Leahy, J.\ P.\ 2001a, \apj, 551, 206
\bibitem[Perlman et al.(2001b)]{perlman01b}
	Perlman, E.\ S., Sparks, W.\ B., Radomski, J., Packham, C.,
	Biretta, J.\ A., and Fisher, R.\ S.\ 2001b, \apjl, submitted
\bibitem[Perlman et al.(1999)]{perlman99}
	Perlman, E.\ S.,  Biretta, J.\ A., Zhou, F., Sparks, W.\ B.,
	and Macchetto, F.\ D. 1999, \aj, 117, 2185
\bibitem[Reynolds et al.(1996)]{reynolds96}
	Reynolds, C.\ S., Di Matteo, T., Fabian, A.\ C., Hwang, U.,
	\& Canizares, C.\ R.\ 1996, \mnras, 283, L111
\bibitem[Schreier et al.(1980)]{schreier82}
	Schreier, E.\ J., Gorenstein, P., and
	Feigelson, E.\ D. 1980, \apj, 261, 42
\bibitem[Schwartz et al.(2000)]{schwartz00}
	Schwartz et al. 2000, \apjl, 540, L69
\bibitem[Sparks et al.(1996)]{sbm96}
	Sparks, W.\ B., Biretta, J.\ A., and Macchetto, F.\ 1996, \apj, 473, 254
\bibitem[Tonry (1991)]{tonry91}
	Tonry, J.\ L. 1991, \apjl, 373, L1
\end{thebibliography}
\end{document}